\documentclass{article}

\usepackage{microtype}
\usepackage{graphicx}
\usepackage{multicol}
\usepackage{subfigure}
\usepackage{makecell}
\usepackage{booktabs} 
\usepackage{multirow} 
\usepackage{hyperref}
\usepackage{amsmath}

\usepackage[accepted]{icml2020}

\icmltitlerunning{EfficientTTS: An Efficient and High-Quality Text-to-Speech Architecture}

\UseRawInputEncoding
\begin{document}

\twocolumn[
\icmltitle{EfficientTTS: An Efficient and High-Quality Text-to-Speech Architecture}

\begin{icmlauthorlist}
    \icmlauthor{Chenfeng Miao}{pa}
    \icmlauthor{Shuang Liang}{pa}
    \icmlauthor{Zhengchen Liu}{pa}
    \icmlauthor{Minchuan Chen}{pa}
    \icmlauthor{Jun Ma}{pa}
    \icmlauthor{Shaojun Wang}{pa}
    \icmlauthor{Jing Xiao}{pa}
\end{icmlauthorlist}

\icmlkeywords{Non-autoregressive model, Text-to-Speech, Monotonic alignments }

\icmlaffiliation{pa}{Ping An Technology}
\icmlcorrespondingauthor{Chenfeng Miao}{miao\_chenfeng@126.com}

\vskip 0.3in
]

\printAffiliationsAndNotice{}

\begin{abstract}


In this work, we address the Text-to-Speech (TTS) task by proposing a non-autoregressive architecture called EfficientTTS. Unlike the dominant non-autoregressive TTS models, which are trained with the need of external aligners, EfficientTTS optimizes all its parameters with a stable, end-to-end training procedure, while allowing for synthesizing high quality speech in a fast and efficient manner. EfficientTTS is motivated by a new monotonic alignment modeling approach (also introduced in this work), which specifies monotonic constraints to the sequence alignment with almost no increase of computation. By combining EfficientTTS with different feed-forward network structures, we develop a family of TTS models, including both text-to-melspectrogram and text-to-waveform networks. We experimentally show that the proposed models significantly outperform counterpart models such as Tacotron 2 \cite{Tacotron2} and Glow-TTS \cite{Glow-TTS} in terms of speech quality, training efficiency and synthesis speed, while still producing the speeches of strong robustness and great diversity. In addition, we demonstrate that proposed approach can be easily extended to autoregressive models such as Tacotron 2.  \footnote{Audio samples of the proposed models are available at: \url{https://mcf330.github.io/EfficientTTS/}}

\end{abstract}

\section{Introduction}

Text-to-Speech (TTS) is an important task in the speech processing. With rapid progress in deep learning, TTS technology has received widespread attention in recent years. The most popular neural TTS models are autoregressive models based on an encoder-decoder framework \cite{Tacotron,Tacotron2,deepvoice3,clarinet,transformerTTS,Flowtron}. In this framework, the encoder takes the text sequence as input and learns its hidden representation, while the decoder generates the outputs frame by frame, i.e., in an autoregressive manner. As the performance of autoregressive models has been substantially promoted, the synthesis efficiency is becoming a new research hotspot. 

Recently, significant efforts have been dedicated to the development of non-autoregressive TTS models \cite{fastspeech,fastspeech2,flow-tts,ParaNet}. However, most existing non-autoregressive TTS models suffer from complex training procedures, high computational cost or training time cost, making them not suited for real-world applications. In this work, we propose EfficientTTS, an efficient and high-quality text-to-speech architecture. Our contributions are summarized as follows,
\begin{itemize}
    \setlength{\itemsep}{0pt}
\item We propose a novel approach to produce soft or hard monotonic alignments for sequence-to-sequence models in addition to a general attention mechanism with almost no increase in computation. Most important, proposed approach can be incorporated into any attention mechanisms without constraints on network structures.
\item We propose EfficientTTS, a non-autoregressive architecture to perform high-quality speech generation from text sequence without additional aligners. EfficientTTS is fully parallel, fully convolutional, and is trained end-to-end, thus being quite efficient for both training and inference.
\item  We develop a family of TTS models based on EfficientTTS, including: (1) EFTS-CNN, a convolutional model learns melspectrogram with high training efficiency; (2) EFTS-Flow, a flow-based model enable parallel melspectrogram generation with controllable speech variation; (3) EFTS-Wav, a fully end-to-end model directly learns waveform generation from text sequence. We experimentally show that proposed models achieve significant improvements in speech quality, synthesis speed and training efficiency, in comparison with counterpart models Tacotron 2 and Glow-TTS.
\item  We also show that proposed approach can be easily extended to autoregressive models such as Tacotron 2 at the end of this paper.
\end{itemize}

The rest of the paper is structured as follows. Section $2$ discusses related work. We introduce monotonic alignment modeling using \textit{index mapping vector} in Section $3$. The EfficientTTS architecture is introduced in Section $4$. In Section $5$, the EfficientTTS models are presented. Section $6$ demonstrates experimental results and implementation details. Finally, Section $7$ concludes the paper.

\section{Related Work}
\subsection{Non-Autoregressive TTS models} \label{nat}
In TTS tasks, an input text sequence $\boldsymbol{x} = \{x_0,x_1,...,x_{T_1-1}\}$ is transduced to an output sequence $\boldsymbol{y} = \{y_0,y_1,...,y_{T_2-1}\}$ through an encoder-decoder framework \cite{encoder-decoder}.\footnote{Throughout the paper, bold letters represent random variables and non-bold letters for realizations of their corresponding random variables.} Typically, $\boldsymbol{x}$ is first converted to a sequence of hidden states $\boldsymbol{h}= \{ h_0,h_1,...,h_{T_1-1}\}$ through an encoder $f$: $\boldsymbol{h}=f(\boldsymbol{x})$, and then passed through a decoder to produce the output $\boldsymbol{y}$. For each output timestep, an attention mechanism allows for searching the whole elements of $\boldsymbol{h}$ to generate a context vector $\boldsymbol{c}$:
\begin{equation}
c_{j} = \sum_{i=0}^{T_1-1} \alpha_{i,j}*h_i, \label{mapping_alpha}
\end{equation}
where $\boldsymbol{\alpha} = \{ \alpha_{i,j}\}\in {\mathcal{R}^{(T_1,T_2)}}$ is the alignment matrix. $\boldsymbol{c}$ is then fed to another network $g$ to generate the output $\boldsymbol{y}$: $\boldsymbol{y}=g(\boldsymbol{c})$. Networks of $f$ and $g$ could be easily replaced with parallel structures because both of them obtain consistent lengths of input and output. Therefore, the key to build a non-autoregressive TTS model lies on parallel alignment prediction.
In previous works, most non-autoregressive TTS models learn alignments from external models or tools \cite{ParaNet,fastspeech,fastspeech2}, making the training complex.
More recently, Flow-TTS \cite{flow-tts}, Glow-TTS \cite{Glow-TTS} and EATS \cite{EATS} are proposed. Flow-TTS and EATS directly learn the alignments from hidden representations of text sequence during training, without considering extracting the alignments from output sequence, making the training inefficient. Glow-TTS extracts the duration of each input token using an independent algorithm which precludes the use of standard back-propagation. EfficientTTS, on the other hand, jointly learns sequence alignment and speech generation through a single network in a fully end-to-end manner, while maintaining a stable and efficient training.

\subsection{Monotonic alignment modeling}
\begin{figure}
    \centering
    \includegraphics[width=7.56cm,height=5.24cm]{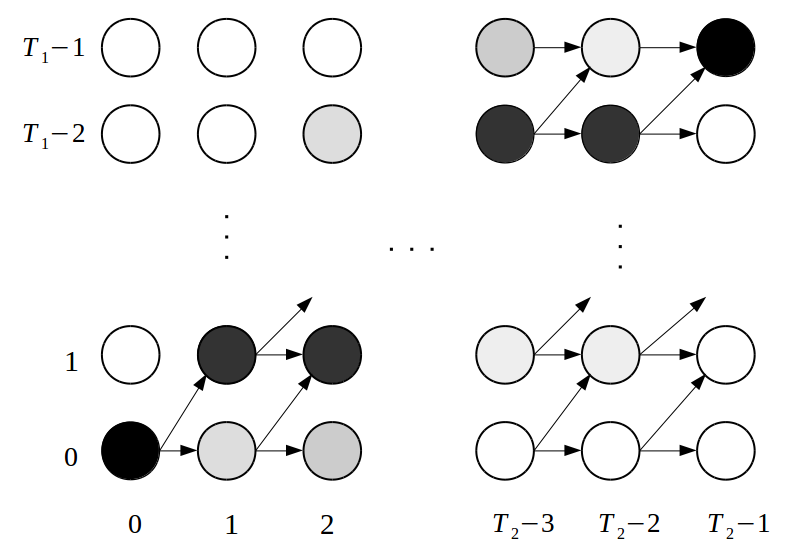}
    \caption{Schematics of the monotonic alignment. Each node $\alpha_{i,j}$ represents the possibility that output timestep $y_j$(horizontal axis) attends on the input token $x_i$(vertical axis). At each output timestep, monotonic attention either move forward to next token or stay unmoved.}\label{fig_mono}
 \end{figure}

 As noted in section \ref{nat}, a general attention mechanism inspects every input step at every output timestep. Such a mechanism often encounters misalignment and is quite costly to train, especially for long sequences. Therefore, it must be helpful if there is some prior knowledge incorporated. In general, the monotonic alignment should follow strict criteria, as shown in Fig. \ref{fig_mono}, which include: (1) \textbf{Monotonicity}, at each output timestep, the aligned position never rewinds; (2) \textbf{Continuity}, at each output timestep, the aligned position move forward at most one step; (3) \textbf{Completeness}, the aligned positions must cover all the positions of input tokens. Lots of prior studies have been proposed to ensure correct alignments \cite{MoBoAligner}, but most of them require sequential steps and often fail to meet all the criteria mentioned above. In this work, we propose a novel approach to produce monotonic attention effectively and efficiently. 

\section{Monotonic Alignment Modeling Using IMV}

We start this section by proposing the \textit{index mapping vector} (IMV), and then we leverage IMV in monotonic alignment modeling. We further show how to incorporate IMV into a general sequence-to-sequence model. 

\subsection{Definition of IMV }

Let $\boldsymbol{\alpha} \in \mathcal{R}^{(T_1,T_2)}$ be the alignment matrix between input sequence $\boldsymbol{x} \in \mathcal{R}^{(D_1,T_1)}$ and output sequence $\boldsymbol{y} \in \mathcal{R}^{(D_2,T_2)}$. We define \textit{index mapping vector} (IMV) $\boldsymbol{\pi}$ as sum of index vector $\boldsymbol{p}=\{0,1,\cdots,T_1-1\}$, weighted by $\boldsymbol{\alpha}$:
\begin{equation}
\pi_{j} = \sum_{i=0}^{T_1-1} \alpha_{i,j}*p_i,  \label{defini}
\end{equation}
where, $0 \le j \le T_2-1$, $\boldsymbol{\pi} \in \mathcal{R}^{T_2}$, and $\displaystyle\sum_{i=0}^{T_1-1} \alpha_{i,j}=1$. We can understand IMV as the \textit{expected location} for each output timestep, where the expectation is over all possible input locations ranging from $0$ to $T_1-1$.

\subsection{Monotonic alignment modeling using IMV}
\textbf{Continuity and Monotonicity.} We first show that the continuity and monotonicity criteria of alignment matrix $\boldsymbol{\alpha}$ is equivalent to the following constraint:
\begin{equation}
    0 \le \Delta \pi_i \le 1  , \label{constriant1}
\end{equation}
where, $\Delta \pi_i = \pi_i - \pi_{i-1}, 1 \le i \le T_2-1$. Detailed verification is shown in Appendix A.

\textbf{Completeness}. Given $\pi$ is continuous and monotonic, completeness is equivalent to boundary conditions:
\begin{align}
    \pi_0 &= 0,     \label{constriant2}       \\    
    \pi_{T_2-1} &= T_1-1. \label{constriant3} 
\end{align}
This can be deduced from $\boldsymbol{\alpha}_0=\{1,0,...,0\}$ and $\boldsymbol{\alpha}_{T_2-1}=\{0,0,...,1\}$, where $\boldsymbol{\alpha}_0=\{ \alpha_{i,j} \mid  0 \le i \le T_1-1, j= 0 \}$ and $\boldsymbol{\alpha}_{T_2-1}=\{ \alpha_{i,j}  \mid 0 \le i \le T_1-1,j= T_2-1\}$.

\subsection{Incorporate IMV into networks}
We propose two strategies to incorporate IMV into sequence-to-sequence networks: Soft Monotonic Alignment (SMA) and Hard Monotonic Alignment (HMA). 

\textbf{Soft Monotonic Alignment (SMA)}.  To let sequence-to-sentence models be trained with the constraints given by Eq. (\ref{constriant1},\ref{constriant2},\ref{constriant3}), a natural idea is to turn these constraints into training objectives. We formulate these constraints as a SMA loss which is computed as:
\begin{align}
    \mathcal{L}_{\rm SMA} &= \lambda_0\Vert \lvert \Delta \pi \rvert - \Delta \pi \Vert_1  \notag \\
    &+ \lambda_1\Vert \lvert \Delta \pi - 1 \rvert + (\Delta \pi -1 )\Vert_1 \notag \\
    &+ \lambda_2\Vert \frac{\pi_0}{T_1-1} \Vert_2  \notag \\
    &+ \lambda_3\Vert \frac{\pi_{T_2-1}}{T_1-1} -1 \Vert_2, \label{smaloss}
\end{align}
where $\Vert \cdot \Vert _1$ and $\Vert \cdot \Vert _2$ are  $L^1$ norm and $L^2$ norm respectively, $\lambda_0,\lambda_1,\lambda_2,\lambda_3$ are positive coefficients. As can be seen, $\mathcal{L}_{\rm SMA}$ is non-negative, and it is zero only if $\boldsymbol{\pi}$ satisfies all the constraints. Computation of $\mathcal{L}_{\rm SMA}$ requires alignment matrix $\boldsymbol{\alpha}$ only (index vector $\boldsymbol{p}$ is always known), therefore, it is quite easy to incorporate SMA loss into sequence-to-sequence networks without changing their network structures. In general, SMA plays a similar role as Guided Attention \cite{guidedattention}. However, SMA outperforms Guided Attention because SMA theoretically provides more accurate constraints on the alignments.

\textbf{Hard Monotonic Alignment (HMA)}.
While SMA allows sequence-to-sequence networks to produce monotonic alignments by incorporating with a SMA loss, the training of these networks may remain costly because the networks cannot produce monotonic alignments at the beginning phase of training. Instead, they learn this ability step by step. To address this limitation, we propose another monotonic strategy which we call HMA, for Hard Monotonic Alignment. The core idea of HMA is to build a new network with a strategically designed structure, allowing for producing monotonic alignments without supervision.

First, we compute IMV $ \boldsymbol{\pi}'$ from the alignment matrix $\boldsymbol{\alpha}$ according to Eq. (\ref{defini}). Although $ \boldsymbol{\pi}'$ is not monotonic, it is then transformed to $\boldsymbol{\pi}$, a strictly monotonic IMV by enforcing $\Delta \boldsymbol{\pi} > 0$ using a $ \rm ReLU$ activation.
\begin{align}
    &\Delta \pi'_j = \pi'_j - \pi'_{j-1}, & 0<j \le T_2-1, \label{step1} \\
    &\Delta \pi_j = \rm ReLU(\Delta \pi'_j), &0<j \le T_2-1, \label{step2}
\end{align}
\begin{equation}
    \pi_j = \begin{cases}
        0, &j= 0\\
        \sum_{m=0}^{j}\Delta {\pi_m}. & 0<j \le T_2-1.
       \end{cases}
       \label{step3}
\end{equation}

Furthermore, to restrict the domain of $\boldsymbol{\pi}$ to the interval $[0,T_1-1]$ as given in Eq. (\ref{constriant2},\ref{constriant3}), we multiply $\boldsymbol{\pi}$ by a positive scalar: 
\begin{align}
    \pi^*_j &= \pi_j * \frac{T_1-1}{\max (\boldsymbol{\pi})} & \notag \\
    &= \pi_j * \frac{T_1-1}{\pi_{T_2-1}}, & 0 \le j \le T_2-1. \label{scale_func}
\end{align}

Recall that our goal is to construct a monotonic alignment. To achieve this, we introduce the following transformation to reconstruct the alignment by leveraging a Gaussian kernel centered on $\boldsymbol{\pi}^*$ :
\begin{equation}
    \alpha'_{i,j} = \frac{\exp{(-\sigma^{-2}(p_i - \pi^*_j)^2)}}{ \sum_{m=0}^{T_1-1}\exp{(-\sigma^{-2}(p_m - \pi^*_j)^2)}}, \label{recons2}
\end{equation}
where, $\sigma^2$ denotes the hyper-parameter representing alignment variation.  $\boldsymbol{\alpha'}$ serves as a replacement of original alignment $\boldsymbol{\alpha}$. The difference between $\boldsymbol{\alpha}$ and $\boldsymbol{\alpha'}$ is that $\boldsymbol{\alpha'}$ is guaranteed to be monotonic while $\boldsymbol{\alpha}$ has no constraint on monotonicity. HMA reduces the difficulty of learning monotonic alignments thus improves the training efficiency. Similar to SMA, HMA can be employed to any sequence-to-sequence networks.

\section{EfficientTTS Architecture}
\begin{figure*}[t]
    \centering
    \subfigure{}{
    \begin{minipage}[t]{.48\linewidth}
    \includegraphics[width=6.0cm]{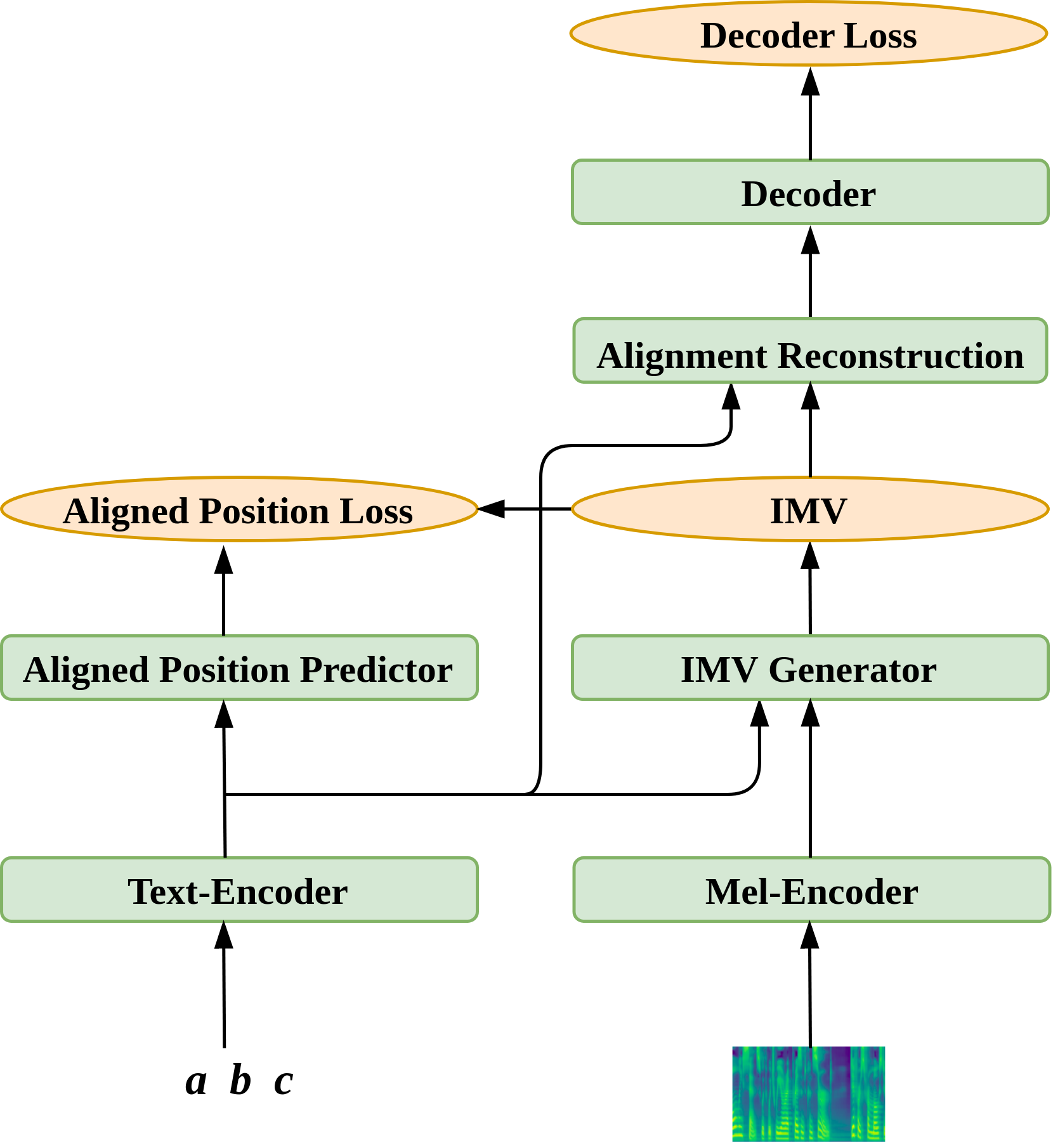}
    \centering
    \centerline{(a) Training phase}
    \end{minipage}
    }
    \subfigure{}{
    \begin{minipage}[t]{.48\linewidth}
    \includegraphics[width=3.5cm]{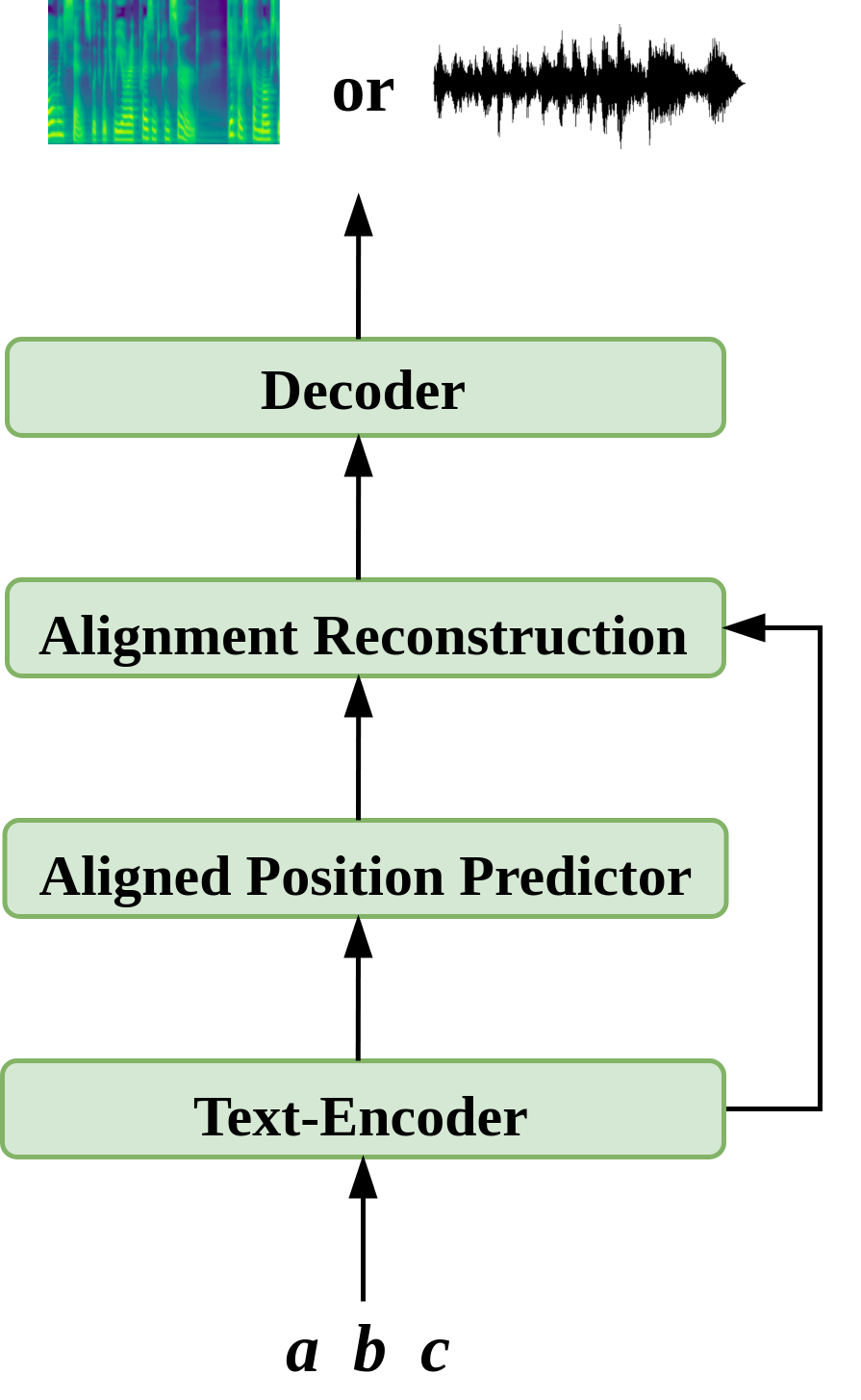}
    \centering
    \centerline{(b) Inference phase}
    \end{minipage}
    }
    \caption{Overall model architecture. }
    \label{model-arch}
  \end{figure*}

The overall architecture design of EfficientTTS is shown in Fig. \ref{model-arch}. In the training phase we learn the IMV from the hidden representations of text sequence and melspectrogram through an IMV generator. The hidden representations of text sequence and melspectrogram are learned from a text-encoder and a mel-encoder respectively. IMV is then converted to a 2-dimensional alignment matrix which is further used to generate the time-aligned representation through an alignment reconstruction layer. The time-aligned representation is passed through a decoder producing the output melspectrogram or waveform. We concurrently train an aligned position predictor which learns to predict aligned position in output timestep for each input text token. In the inference phase, we reconstruct the alignment matrix from predicted aligned positions. We show detailed implementation in the following subsections and pseudocode of each components in Appendix D.
\subsection{Text-Encoder and Mel-Encoder}
We use a text-encoder and a mel-encoder to convert text symbols and melspectrograms to powerful hidden representations respectively.

In the implementation of the text-encoder, we use learned embedding to covert the text sequence to a sequence of high-dimensional vectors. The high-dimensional vectors are then passed through a stack of convolutions interspersed with weight normalization \cite{wn} and Leaky ReLU activation. We also add a residual connection for each convolution to allow for deep networks.

In the implementation of the mel-encoder, we first convert melspectrograms to  high-dimensional vectors through a linear projection. Same as the text-encoder, mel-encoder consists of a stack of convolutions interspersed with weight normalization, Leaky ReLU activation, and residual connection. Note that mel-encoder is only used in the training phase.

\subsection{IMV generator}
In order to generate a monotonic IMV in the training phase, we first learn the alignment $\boldsymbol{\alpha}$ between the input and output through a scaled dot-product attention \cite{Transformer} as given in Eq. (\ref{sa}), and then compute IMV from $\boldsymbol{\alpha}$. 
\begin{equation}
    \alpha_{i,j} = \frac{\exp{(-D^{-0.5}(\boldsymbol{q_j} \cdot \boldsymbol{k_i}))}}{ \sum_{m=0}^{T_1-1}\exp{(-D^{-0.5}(\boldsymbol{q_j} \cdot \boldsymbol{k_m}))}}, \label{sa}
\end{equation}
where, $\boldsymbol{q}$ and $\boldsymbol{k}$ are the outputs of mel-encoder and text-encoder, and $D$ is the dimensionality of $\boldsymbol{q}$ and $\boldsymbol{k}$. 

A simple way to compute IMV is to follow Eq. (\ref{defini}). However, since scaled dot-product attention has no constraint on monotonicity, in our preliminary experiments, a SMA loss was incorporated for training. But we further discovered that HMA is more efficient. We follow Eq. (\ref{step1},\ref{step2},\ref{step3},\ref{scale_func}) in implementing HMA. In experiments we compare the effects of different monotonic strategies.

\subsection{Aligned position predictor}

In the inference phase, the model needs to predict the IMV $\boldsymbol{\pi}$ from the hidden representation of text sequence $\boldsymbol{h}$, which is challenging in practice. There are two limitations: (1) $\boldsymbol{\pi}$ is time-aligned, which is in high resolution but $\boldsymbol{h}$ is in low resolution; (2) Each prediction of $\pi_i$ affects later prediction of $\pi_j$ $(j > i)$ due to cumulative sum operation introduced in Eq. (\ref{step3}), making it difficult to predict $\boldsymbol{\pi}$ in parallel. Fortunately, the limitations can be alleviated by predicting the aligned positions $\boldsymbol{e}$ of each input token instead.
\begin{figure}
    \centering
    \includegraphics[width=7.56cm,height=5.24cm]{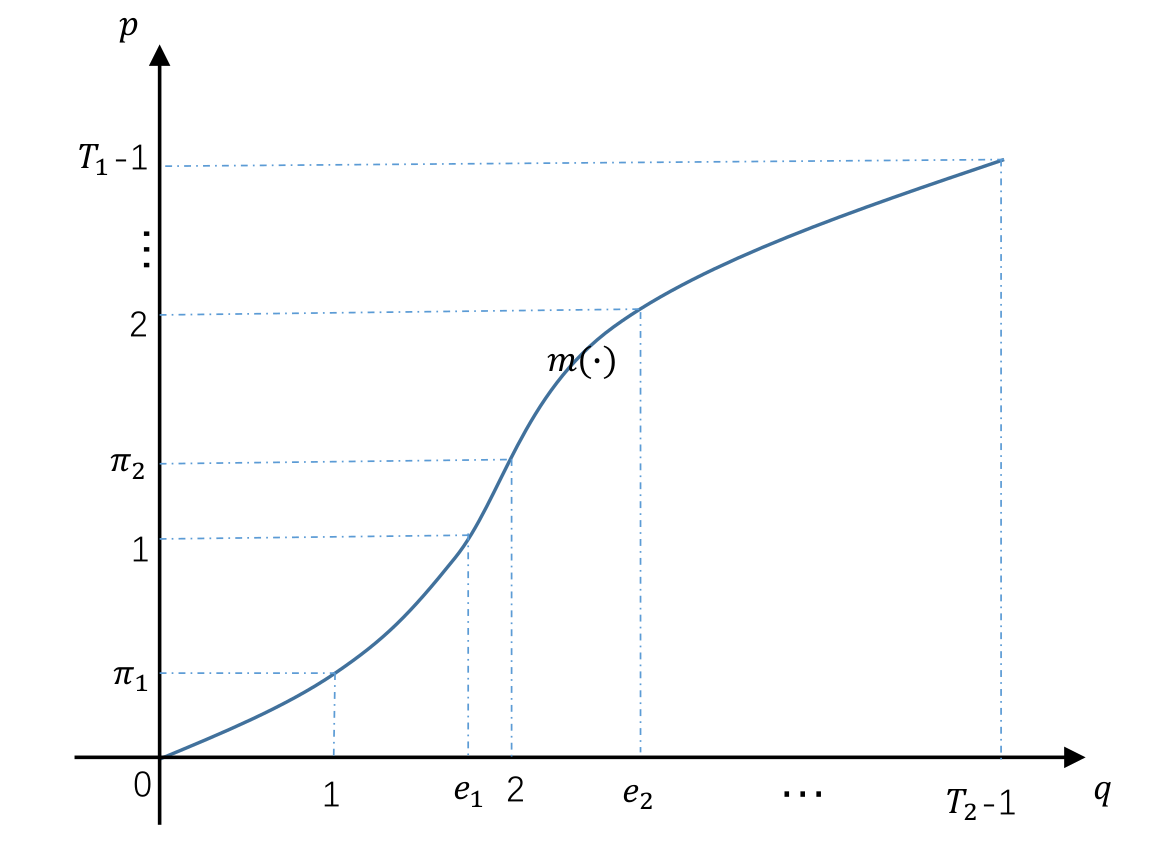}
    \caption{Schematics of $m(\cdot)$. $\boldsymbol{p},\boldsymbol{q}$ are indexes of input sequence and output sequence respectively. $\boldsymbol{\pi}$ is IMV, $\boldsymbol{e}$ is the aligned position in output timestep for each input token.}\label{shi}
\end{figure}
We define Eq. (\ref{defini}) as transformation $m(\cdot)$: $\boldsymbol{\pi} = m(\boldsymbol{q})$. Since both $\boldsymbol{\pi}$ and $\boldsymbol{q} $ are monotonic and continuous across timesteps, which means transformation $m(\cdot)$ is monotonic and continuous, thus $m(\cdot)$ is invertible:
\begin{equation}
    \boldsymbol{q}= m^{-1}(\boldsymbol{\pi}),
\end{equation} 
The aligned positions $\boldsymbol{e}$ in output timestep for each input token can be computed as: 
$$ \boldsymbol{e}= m^{-1}(\boldsymbol{p}),\qquad  \boldsymbol{p}=\{0,1,...,T_1-1\}.$$
We illustrate the relations of $m(\cdot),\boldsymbol{e},\boldsymbol{\pi}$ in Fig. \ref{shi}. 
In order to compute $\boldsymbol{e}$, we first compute the probability density matrix $\boldsymbol{\gamma}$ utilizing a similar transformation as Eq. (\ref{recons2}). The only difference is that the probability density is computed on different dimensions. The aligned position $\boldsymbol{e}$ is the weighted sum of the output index vector $\boldsymbol{q}$ weighted by $\boldsymbol{\gamma}$.
\begin{equation}
    \gamma_{i,j} = \frac{\exp{(-\sigma^{-2}(p_i - \pi_j)^2)}}{ \sum_{n=0}^{T_2-1}\exp{(-\sigma^{-2}(p_i - \pi_n)^2)}}, \label{recons_new}
\end{equation}
\begin{equation}
    e_{i} = \sum_{n=0}^{T_2-1} \gamma_{i,n}*q_n. \label{e_calc}
\end{equation}
As can be seen, the computation of $\boldsymbol{e}$ is differentiable which allows for training by gradients methods, thus can be used in both training and inference. Besides, $\boldsymbol{e}$ is predictable, because:(1) The resolution $\boldsymbol{e}$ is same as $\boldsymbol{h}$; (2) we can learn relative position $\Delta \boldsymbol{e}, (\Delta e_i = e_i- e_{i-1}, 1 \le i \le T_1-1)$ instead of directly learning $\boldsymbol{e}$ to overcome the second limitation.

The aligned position predictor consists of 2 convolutions, each followed by the layer normalization and ReLU activation. We regard $\Delta \boldsymbol{e}$ computed from $\boldsymbol{\pi}$ as the training target. The loss function between the estimated position $\Delta \hat{\boldsymbol{e}}$ and the target one $\Delta \boldsymbol{e}$ is computed as:
\begin{equation}
    \mathcal{L}_{ap} = \Vert \log{(\Delta \hat{\boldsymbol{e}} + \epsilon )} - \log{(\Delta \boldsymbol{e} + \epsilon)} \Vert_1,
\end{equation}
Where, $\epsilon$ is a small number to avoid numerical instabilities. The goal with log-scale loss is to accurately fit small values, which tends to be more important towards the later phases of training. Aligned position predictor is learned jointly with the rest of the model. Because we generate alignments by leveraging the aligned positions, as a side benefit, EfficientTTS inherits the ability of speech rate control as duration-based non-autoregressive TTS models. 

\subsection{Alignment reconstruction}

In order to map input hidden representations $\boldsymbol{h}$ to time-aligned representations, an alignment matrix is needed, for both training and inference. We can alternatively construct alignment from IMV or the aligned positions. For most situations, Eq. (\ref{recons2}) is an effective way to reconstruct alignment matrix from IMV. But because we predict aligned positions rather IMV during inference, to be consistent, we reconstruct alignment matrix from the aligned positions $\boldsymbol{e}$ for both training and inference. Specifically, we take the aligned positions $\boldsymbol{e}$ computed from Eq. (\ref{e_calc}) for training, and the predicted one from aligned position predictor for inference.

We follow similar idea of EATS \cite{EATS} in reconstructing the alignment matrix $\boldsymbol{\alpha'}$ by introducing a Gaussian kernel centered on aligned position $\boldsymbol{e}$.
\begin{equation}
    \alpha'_{i,j} = \frac{\exp{(-\sigma^{-2}(e_i - q_j)^2)}}{ \sum_{m=0}^{T_1-1}\exp{(-\sigma^{-2}(e_m - q_j)^2)}},
\end{equation}
where $\boldsymbol{q}=\{0,1,...,T_2-1\}$ is the index vector of output sequence. The length of output sequence $T_2$ is known in training and computed from $\boldsymbol{e}$ in inference:
\begin{equation}
    T_2 = e_{T_1-1} + \Delta e_{T_1-1}.
\end{equation}
Although the reconstructed alignment matrix maybe not as accurate as the one computed by Eq. (\ref{recons2}) (due to the low resolution of $\boldsymbol{e}$), the effect on the output is small because the network is able to compensate. As a result, we enjoy improvement in speech quality caused by the increasing consistency of training and inference. 

\begin{figure*}[t]
    \includegraphics[width=16.5cm,height=4.8cm]{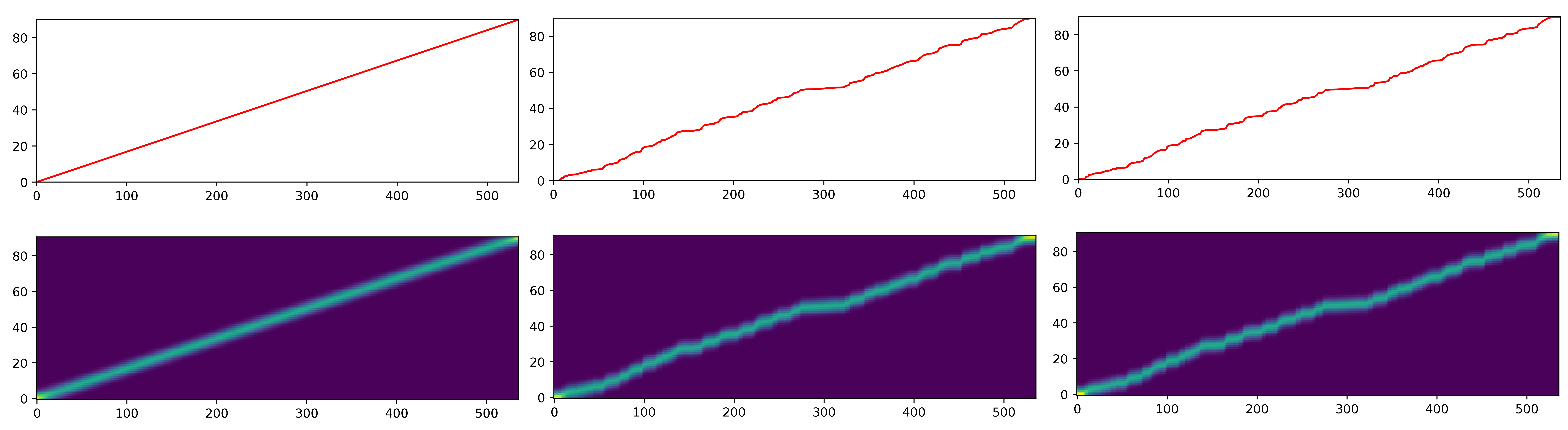}
    \caption{ IMV and reconstructed alignments in training phase of different training steps. The training steps are $1$, $10$k and $270$k respectively. The whole model converges at training step $270$k.}\label{imvpng}
\end{figure*}

We illustrate $\boldsymbol{\pi}$ and the reconstructed $\boldsymbol{\alpha}$ of same utterance in Fig. \ref{imvpng}. As can be seen, $\boldsymbol{\alpha}$ is diagonal at first training step, and it quickly converges at $10$k$^{th}$ training step, which is significantly fast. We map the output of text-encoder $\boldsymbol{h}$ to a time-aligned representation by making use of $\boldsymbol{\alpha}'$ following Eq. (\ref{mapping_alpha}). The time-aligned representation is then fed as input to decoder.

\subsection{Decoder}
Since both of the input and output of decoder are time-aligned, 
it is easy to implement decoder with parallel structures. In next section, we develop three models based on EfficientTTS with different decoder implementations.  

\section{EfficientTTS Models}

\subsection{EFTS-CNN}
We first parameterize the decoder by a stack of convolutions. Each convolution is interspersed with weight normalization, Leaky ReLU activation, and residual connection. We add a linear projection at the end to generate melspectrogram. Mean square error (MSE) is used as the reconstruction error. The overall training objective of EFTS-CNN is a combination of aligned position loss and MSE loss of melspectrogram.

\subsection{EFTS-Flow}
To let TTS model have the ability to control the variations of generated speech, we implement a flow-based decoder. In the training phase, we learn a transformation $f$ from melspectrogram to a high dimensional Gaussian distribution $\mathcal{N}(\boldsymbol{0},\boldsymbol{1})$ by directly maximizing the likelihood, conditioning on the time-aligned representation. $f$ is invertible with a strategically designed structure. Specifically, it consists of several flow steps, and each flow step consists of two elemental invertible transformations: an invertible linear layer and an affine coupling layer. To improve the diversity of generated speech, we sample the latent variable $\boldsymbol{z}$ from  Gaussian distribution $\mathcal{N}(\boldsymbol{0},\boldsymbol{1})$ during inference, and interpret $\boldsymbol{z}$ with a zero vector $\boldsymbol{o}$ using temperature factor $t$, and inverse the transformation $f$ to produce the melspectrogram. 
\begin{align}
    &\boldsymbol{z}' = t*\boldsymbol{z} + \boldsymbol{o}*(1-t),0 \le t \le 1 \\
    &\boldsymbol{x} = f^{-1}(\boldsymbol{z}').
\end{align}
For sake of simplicity we follow the decoder structure of Flow-TTS \cite{flow-tts} in implementing our flow-based decoder.  The overall training objective of EFTS-Flow is a combination of aligned position loss and maximum likelihood estimation (MLE) loss.

\subsection{EFTS-Wav}
To simplify the 2-staged training pipeline and train TTS models in a fully end-to-end manner, we develop a text-to-wav model by incorporating EfficientTTS with a dilated convolutional adversarial decoder. The decoder structure is similar to MelGAN \cite{Melgan} except: (1) The input of the generator is high dimensional hidden representations not 80-channel melspectrograms; (2) A multi-resolution STFT loss is incorporated at end of the generator. We adopt the same structure of MelGAN discriminator for adversarial training. Similar as ClariNet \cite{clarinet} and EATS \cite{EATS}, we train MelGAN part by conditioning on sliced input corresponding to $1$s audio clips, while other part is trained on the whole-length utterance. We add a linear projection on the whole-length decoder input to generate melspectrogram concurrently, which allows EFTS-Wav to learn the whole-length alignment for each training step. The overall training objective of EFTS-Wav generator is a linear combination of reconstruction loss of melspectrogram, MelGAN generator loss, multi-resolution STFT loss, and aligned position loss. 

\section{Experiments}

\begin{table*}[t]
\caption{Quantitative results of training time and inference latency. We run training and inference on a single V100 GPU. We select 20 sentence for inference speed evaluation and run inference of each sentence 20 times to get a average inference latency. The lengths of the generated melspectrograms range from 110 to 802, with an average of 531. We exclude the time cost of transferring data between CPU and GPU in inference speed evaluation. HiFi-GAN is used to produce waveform from melspectrograms.}
\label{speed_table}
\vskip 0.15in
\begin{center}
\begin{small}
\begin{tabular}{lcccccc}
\toprule
Model family & \makecell[c]{Training \\ Time(h)} & \makecell[c]{Training \\ Speedup} & \makecell[c]{Inference Time \\ text-to-mel(ms)} & \makecell[c]{Inference Speedup \\ text-to-mel} & \makecell[c]{Inference Time\\ text-to-wav(ms) } & \makecell[c]{Inference Speedup \\ text-to-wav}\\
\midrule
Tacotron 2 &54  & -  & 780 &-  & 824& - \\
Glow-TTS & 120 & 0.45$\times$ & 42  &18.6$\times$& 86& 9.6$\times$\\
\midrule
EFTS-CNN & 27  & \textbf{2}$\times$ & \textbf{6}  & \textbf{130.0}$\times$ & 50& 16.5$\times$\\
EFTS-Flow & 40 & 1.35$\times$ & 11 &70.9$\times$ & 55& 14.9$\times$\\
EFTS-Wav & -   & -  & -  & - & \textbf{16}& \textbf{54.0}$\times$\\
\bottomrule
\end{tabular}
\end{small}
\end{center}
\vskip -0.1in
\end{table*}
In this section, we first compare proposed models with their counterparts in terms of speech fidelity, training and inference efficiency. We then analyze the effectiveness of proposed monotonic approach on both EFTS-CNN and Tacotron 2. We also demonstrate that proposed models can generate speech in great diversity at the end of this section. 

\subsection{Experimental setup}
\textbf{Datasets}. We conduct most of our experiments on an open-source standard Mandarin dataset from DataBaker\footnote{\url{https://www.data-baker.com/open\_source.html}}, which consists of $10,000$ Chinese clips from a single female speaker with a sampling rate of 22.05kHZ. The length of the clips varies from 1 to 10 seconds and the clips have a total length of about 12 hours. We follow \cite{Tacotron} in converting the waveforms to 80-channel melspectrogram. The FFT size is 1024, hop length is 256, and window size is 1024. We also conduct some experiments using LJ-Speech dataset \cite{ljspeech}, which is a 24-hour waveform audio set of a single female speaker with 131,00 audio clips and a sample rate of 22.05kHZ.

\textbf{Implementation details}. Our implementation of EfficientTTS consists of 5 convolutions in text-encoder and 3 convolutions in mel-encoder, the kernel size and dimension size of all the convolutions are set to 5 and 512 respectively. We use 6-layer convolution stack in EFTS-CNN decoder with the same convolution configurations. The decoder of EFTS-Flow consists of 8 flow steps, we early output 20 channels for every 3 flow steps in implementing the multi-scale architecture. We follow the configurations of MelGAN in implementing of EFTS-wav. We use HiFi-GAN \cite{HiFi-GAN} vocoder to produce waveforms from melspectrograms generated by EFTS-CNN and EFTS-Flow. We use the open implementation of HiFi-GAN\footnote{\url{https://github.com/jik876/hifi-gan}} with HiFi-GAN-V1 configuration.

\textbf{Counterpart models}. We compare proposed models with autoregressive Tacotron 2 and non-autoregressive Glow-TTS in the following experiments. We directly use the open-source implementations of Tacotron 2\footnote{\url{https://github.com/NVIDIA/tacotron2}}
and Glow-TTS\footnote{\url{https://github.com/jaywalnut310/glow-tts}} with default configurations.

\textbf{Training}. We train all models on a single Tesla V100 GPU. For EFTS-CNN and EFTS-Flow, we use the Adam optimizer \cite{adam} with a batch size of 96 and a constant learning rate of $1 \times 10^{-4}$. For EFTS-Wav, we use Adam optimizer with a batch size of 48. We use a learning rate of $1 \times 10^{-4}$ for EFTS-Wav generator and $5 \times 10^{-5}$ for EFTS-Wav discriminator. It takes 270k steps for training EFTS-CNN on DataBaker until converge, and 400k training steps for EFTS-Flow, 560k training steps for EFTS-Wav.

\subsection{Comparison with counterpart models}

\textbf{Speech quality}. We conduct a 5-scale mean opinion score (MOS) evaluation on DataBaker dataset to measure the quality of synthesized audios. Each audio is listened by at least 15 testers, who are all native speakers. We compare the MOS of the audio samples generated by EfficientTTS families with ground truth audios, as well as audio samples generated by counterpart models. The MOS result with $95\%$ confidence intervals is shown in Tab. \ref{MOS_table_biaobei}. We draw the observation that EfficientTTS families outperform counterpart models. Tacotron 2 suffers from a declining speech quality caused by the inconsistency between teacher forcing training and autoregressive inference, and Glow-TTS replicates the hidden representations of text sequence, which corrupts the continuity of hidden representations. EfficientTTS reconstructs the alignments using IMV, which is more expressive than token duration, therefore achieves better speech quality. In addition, the alignment part of EfficientTTS is trained together with the rest of the model, which further improves the speech quality. As our training settings may be different with original settings for counterpart models, we further compare our model EFTS-CNN with pertained models of Tacotron 2 and Glow-TTS on LJ-Speech dataset. As shown in Tab. \ref{MOS_table_lj-speech}, EFTS-CNN significantly outperforms counterpart models on LJ-Speech as well.

\begin{table}[t]
\caption{The MOS with $95\%$ confidence intervals for different methods on DataBaker. The temperature of latent variable $z$ is set to 0.667 for both Glow-TTS and EFTS-Flow.}
\label{MOS_table_biaobei}
\begin{center}
\begin{small}
\begin{tabular}{lc}
\toprule
Method & MOS  \\
\midrule
GT & 4.64 $\pm$ 0.07 \\
GT(Mel+HiFi-GAN) & 4.58 $\pm$ 0.13  \\
\midrule
Tacotron 2(Mel+HiFi-GAN) & 4.20 $\pm$ 0.11 \\
Glow-TTS(Mel+HiFi-GAN) & 3.97 $\pm$ 0.21 \\
\midrule
EFTS-CNN(Mel+HiFi-GAN) & \textbf{4.41  $\pm$ 0.13} \\
EFTS-Flow(Mel+HiFi-GAN) & \textbf{4.35 $\pm$ 0.17}  \\
EFTS-Wav & \textbf{4.40 $\pm$ 0.21}  \\
\bottomrule
\end{tabular}
\end{small}
\end{center}
\vskip -0.1in
\end{table}

\begin{table}[t]
    \caption{The MOS with $95\%$ confidence intervals for different methods on LJ-Speech. The temperature of latent variable $z$ is set to 0.667 for Glow-TTS.}
    \label{MOS_table_lj-speech}
    \begin{center}
    \begin{small}
    \begin{tabular}{lc}
    \toprule
    Method & MOS  \\
    \midrule
    GT & 4.75 $\pm$ 0.12 \\
    GT(Mel+HiFi-GAN) & 4.51 $\pm$ 0.13  \\
    \midrule
    Tacotron 2(Mel+HiFi-GAN) & 4.08 $\pm$ 0.13 \\
    Glow-TTS(Mel+HiFi-GAN) & 4.13 $\pm$ 0.18 \\
    \midrule
    EFTS-CNN(Mel+HiFi-GAN) & \textbf{4.37  $\pm$ 0.13} \\
    \bottomrule
    \end{tabular}
    \end{small}
    \end{center}
    \end{table}

\textbf{Training and Inference speed}. Being non-autoregressive and fully convolutional, proposed models are very efficient for both training and inference. Quantitative results of training time and inference latency are shown in Tab. \ref{speed_table}. As can be seen, EFTS-CNN requires the least amount of training time. Although EFTS-Flow requires comparable training time with Tacotron 2, it is significantly faster than Glow-TTS. As for inference latency, EfficientTTS models are faster than Tacotron 2 and Glow-TTS. In particular, the inference latency of EFTS-CNN is $6$ms which is $130\times$ faster than Tacotron 2, and significantly faster than Glow-TTS. Thanks to the removal of melspectrogram generation, EFTS-Wav significantly faster than 2-staged models, taking only $16$ms to synthesize test audios from text sequences, which is $54\times$ faster than Tacotron 2.  

\subsection{Evaluation of monotonic alignments approach}
In order to evaluate the behaviour of proposed monotonic approach, we conduct several experiments on EFTS-CNN and Tacotron 2. We first compare the training efficiency on EFTS-CNN, and then conduct a robustness test on Tacotron 2 and EFTS-CNN.

\begin{figure}   
    \centering
    \includegraphics[width=7cm,height=5cm]{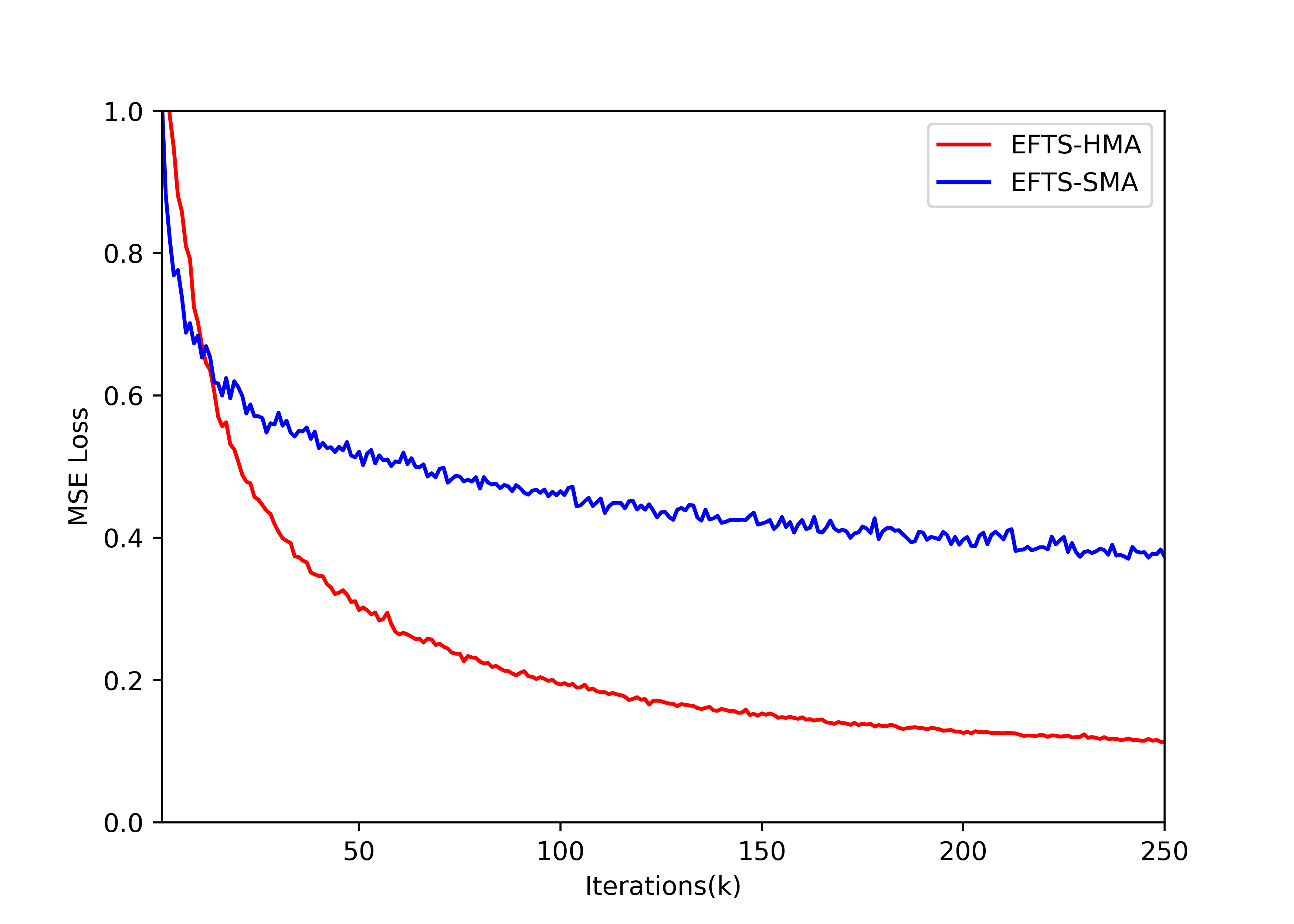}
    \caption{ Training loss of EFTS-HMA and EFTS-SMA.}\label{loss_curve}
\end{figure}

\textbf{Experiments on EFTS-CNN}. We train EFTS-CNN with different settings, including: (1) EFTS-HMA, default implementation of EFTS-CNN, with a hard monotonic IMV generator. (2) EFTS-SMA, an EFTS-CNN model with a soft monotonic IMV generator. (3) EFTS-NM, an EFTS-CNN model with no constant on monotonicity. The network structure of EFTS-NM is same as EFTS-SMA, except that EFTS-SMA is trained with SMA loss while EFTS-NM is trained without SMA loss. We first found that EFTS-NM does not converge at all, its alignment matrix is not diagonal, while both EFTS-SMA and EFTS-HMA are able to produce reasonable alignment. We plot the curves of melspectrogram loss in Fig. \ref{loss_curve} for EFTS-SMA and EFTS-HMA. As can be seen, EFTS-HMA achieves a significant speed-up over EFTS-SMA. Therefore, we can conclude that monotonic alignment is quite essential for proposed models. Our approaches, for both SMA and HMA, succeed to learn monotonic alignments while the vanilla attention mechanism fails. And EFTS-HMA significantly improve the model performance thanks to the strictly monotonic alignment. More training details is shown on Tab. \ref{monotoic_tab}.
\begin{table}[t]
    \caption{Comparison of different monotonic approaches on EFTS-CNN. EFTS-NM does not converge with more than $500k$ training steps. Although EFTS-SMA converges at $450k$th training step, the generated melspectrogram is noisier than EFTS-HMA.}
    \label{monotoic_tab}
    \begin{center}
    \begin{small}
    \begin{tabular}{lcc}
    \toprule
    Models & Training Steps & MSE Loss \\
    \midrule
    EFTS-HMA & 270k & 0.095 \\
    EFTS-SMA & 450k & 0.33\\
    EFTS-NM & not converge &  -\\
    \bottomrule
    \end{tabular}
    \end{small}
    \end{center}
\end{table}
\begin{table}[t]
    \caption{The comparison of robustness between EFTS-CNN and Tacotron 2. We implement Tacotron 2 with different settings, including vanilla Tacotron 2 (T2), Tacotron 2 with SMA (T2-SMA), Tacotron 2 with HMA (T2-HMA). }
    \label{Robustness_table}
    \vskip 0.15in
    \begin{center}
    \begin{small}
    \begin{tabular}{l|ccccc}
    \toprule
    Models & Repeats & Skips & \makecell[c]{Mispron-\\unciations}& \makecell[c]{Error \\ Rate}\\ 
    \midrule
    T2 & 13 & 7 &  5  & 50$\%$\\
    T2-SMA & 3 & 1 & 3  & 14$\%$\\
    T2-HMA & 0 & 1 & 3 & 8$\%$ \\
    EFTS-CNN & 0 & 0 & 2 &  4$\%$ \\
    \bottomrule
    \end{tabular}
    \end{small}
    \end{center}
\end{table}
 
\textbf{Robustness}. Many TTS models encounter misalignment at synthesis, especially for autoregressive models. We analyze the attention errors for EfficientTTS in this subsection, the errors are including: repeated words, skipped words and mispronunciations. We perform a robustness evaluation on a 50-sentence test set, which includes particularly challenging cases for TTS systems, such as particularly long sentences, repeated letters etc. We compare EFTS-CNN with Tacotron 2. We also incorporate SMA and HMA into Tacotron 2 for a more detailed comparison (The detailed implementations of Tacotron2-SMA and Tacotron2-HMA and more experimental results are shown in appendix C.). The experimental results of robustness test are shown in Tab. \ref{Robustness_table}. It can be seen that EFTS-CNN effectively eliminate repeats errors and skips errors while Tacotron 2 encounters many errors. However, the synthesis errors are significantly reduced for Tacotron 2 by leveraging SMA or HMA, which indicates that proposed monotonic approach can improve the robustness for TTS models.

\subsection{Diversity}
To synthesize speech samples in great diversity, most of TTS models make use of external conditions such as style embedding or speaker embedding, or just rely on drop-out during inference. However, EfficientTTS is able to synthesize varieties of speech samples in several ways, including:
(1) Synthesizing speech with different alignment scheme. The alignment scheme could either be an IMV which is extracted from existing audio by mel-encoder and IMV generator, or a sequence of duration or a sequence of aligned positions;
(2) Synthesizing speech with different speech rate by multiplying a scalar across predicted aligned positions, which is similar to other duration-based non-autoregressive models;
(3) Synthesizing speech with different speech variations for EFTS-Flow by changing the temperature $t$ of latent variable $z$ during inference. We plot varieties of melspectrograms generated from the same text sequences in appendix B.  

\section{Conclusions And Future Works}

It is often assumed that there exists an unavoidable trade-off between model efficiency versus speech quality. Autoregressive models, such as Tacotron 2 and TransformerTTS, achieve human-like speech quality but generally suffer from slow synthesis speed due to their autoregressive structure. Non-autoregressive models can synthesize quickly, but cannot be trained efficiently. In this work, we propose a non-autoregressive architecture which enables high quality speech generation as well as efficient training and synthesis. We develop a family of models based on EfficientTTS covering text-to-melspectrogram and text-to-waveform generation. Through extensive experiments, we observe improved quantitative results including training efficiency, synthesis speed, robustness, as well as speech quality. We show that proposed models are very competitive compared with existing TTS models.  

There are many possible directions for future work. EfficientTTS enables not only generating speech at a given alignment and but also extracting alignment from given speech, making it an excellent candidate for voice conversion and singing synthesis. It is also a good choice to apply the proposed monotonic approach to other sequence-to-sequence tasks where monotonic alignment matters, such as Automatic Speech Recognition (ASR), Neural Machine Translation (NMT), and Optical Character Recognition (OCR). Besides, we are also very interested in further investigations on IMV, including its strengths and weaknesses in comparison with an alignment matrix.

\bibliography{example_paper}
\bibliographystyle{icml2020}
\clearpage
\onecolumn 
    
\appendix
\section{Verification of Eq. (\ref{constriant1}) }
In this section, we verify that the continuity and monotonicity of alignment $\boldsymbol{\alpha}$ is equivalent to $0 \le \Delta \boldsymbol(\pi) \le 1$.

We denote $P(y_j,x_i)$ the probability that the decoder frame $y_i$ is attended on the text token $x_i$. At each output timestep, an alignment pair $(y_i,x_i)$ either move forward by one step to $(y_{j+1}, x_{i+1})$ or stay unmoved $(y_{j+1}, x_{i})$. 
Then we have:
\begin{equation}
  P(y_{j+1},x_{i+1}|y_j,x_i) +  P(y_{j+1},x_i|y_j,x_i) = 1, \notag
\end{equation}
where $P(y_{j+1},x_{i+1}|y_j,x_i)$ is the conditional probability that $(y_{j}, x_{i})$ move forward to $(y_{j+1}, x_{i+1})$ given $y_i$ is attended on $x_i$. 

For convenience we define:
 $$ \beta_{i,j} = P(y_{j+1},x_i|y_j,x_i),$$
 $$\alpha_{i,j} = P(y_j,x_i).$$
Thus we have:
\begin{align}
    \alpha_{i,j} &= P(y_j,x_i | y_{j-1},x_i)*P(y_{j-1},x_i) + P(y_j,x_i | y_{j-1},x_{i-1})*P(y_{j-1},x_{i-1}) \notag \\
    &= P(y_j,x_i | y_{j-1},x_i)*P(y_{j-1},x_i) +(1- 
    P(y_j,x_{i-1}) | y_{j-1},x_{i-1}))*P(y_{j-1},x_{i-1}) \notag \\
    & = \beta_{i,j-1}*\alpha_{i,j-1}
 + (1 - \beta_{i-1,j-1})*\alpha_{i-1,j-1}. 
\end{align}

And further:
\begin{align}
    \pi_{j} &= \sum_{i=0}^{T_1-1} \alpha_{i,j}*p_i  \notag \\
            &= \sum_{i=0}^{T_1-1}\beta_{i,j-1}*\alpha_{i,j-1}  *p_i  +  \sum_{i=1}^{T_1-1}(1-\beta_{i-1,j-1})*\alpha_{i-1,j-1} * p_i.
\end{align}

Because $p_i = p_{i-1} + 1, \forall 1 \le i \le T_1-1$, we have:

\begin{align}
    \pi_{j}  =& \sum_{i=0}^{T_1-1}\beta_{i,j-1}*\alpha_{i,j-1}  *p_i +  \sum_{i=1}^{T_1-1}(1-\beta_{i-1,j-1})*\alpha_{i-1,j-1} * (p_{i-1} + 1)  \notag \\
    =& \sum_{i=0}^{T_1-1}\beta_{i,j-1}*\alpha_{i,j-1}*p_i + \sum_{i=1}^{T_1-1} \alpha_{i-1,j-1}*p_{i-1}  \notag \\
    -& \sum_{i=1}^{T_1-1}\beta_{i-1,j-1}*\alpha_{i-1,j-1}*p_{i-1} + \sum_{i=1}^{T_1-1}(1-\beta_{i-1,j-1})*\alpha_{i-1,j-1}  \notag \\
    =& \pi_{j-1} + \sum_{i=1}^{T_1-1}(1-\beta_{i-1,j-1})*\alpha_{i-1,j-1}.
    \end{align}

Because:
\begin{equation}
0 \le \sum_{i=1}^{T_1-1}(1-\beta_{i-1,j-1})*\alpha_{i-1,j-1} \le \sum_{i=0}^{T_1-1}\alpha_{i,j-1} =1.
\end{equation}
Then we get:
\begin{equation}
    0 \le \Delta \pi_j \le 1,\forall j \in [0,T_2-1].
\end{equation}

\section{Analysis of Melspectrograms}

\begin{figure}   
    \centering
    \includegraphics[width=7.1cm,height=6.9cm]{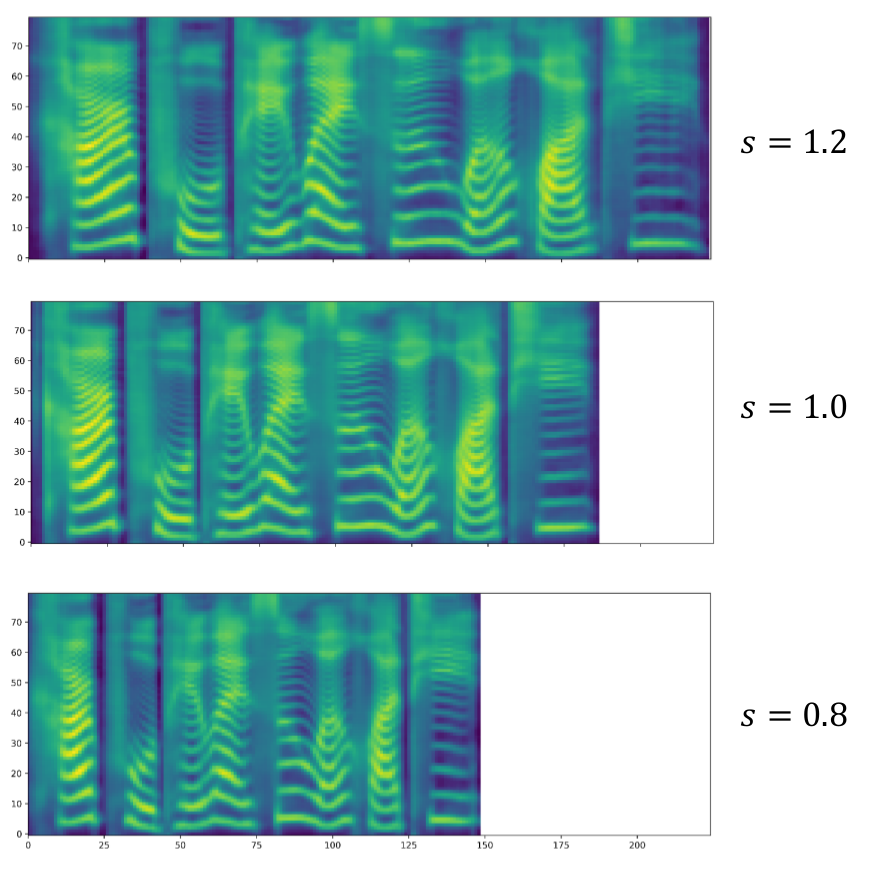}
    \caption{ Melspectrograms with different speech rate by multiplying the aligned position by a positive scalar.}\label{speech_rate}
 \end{figure}

\textbf{Speech rate}.
EfficientTTS is able to produce high quality speech with different speech rate by multiplying the aligned position with a positive scalar. Fig. \ref{speech_rate} shows melspectrograms with different speech rate of the same utterance. The scalar factors are $1.2,1.0,0.8$ respectively. As can be seen, the generated melspectrograms are similar with each other, which means proposed models are able to speed up or slow down generated speech without changing the pitch and degrading the speech quality. We also attach several audio samples in our demo page.
 
\begin{figure}   
    \centering
    \includegraphics[width=7.1cm,height=6.9cm]{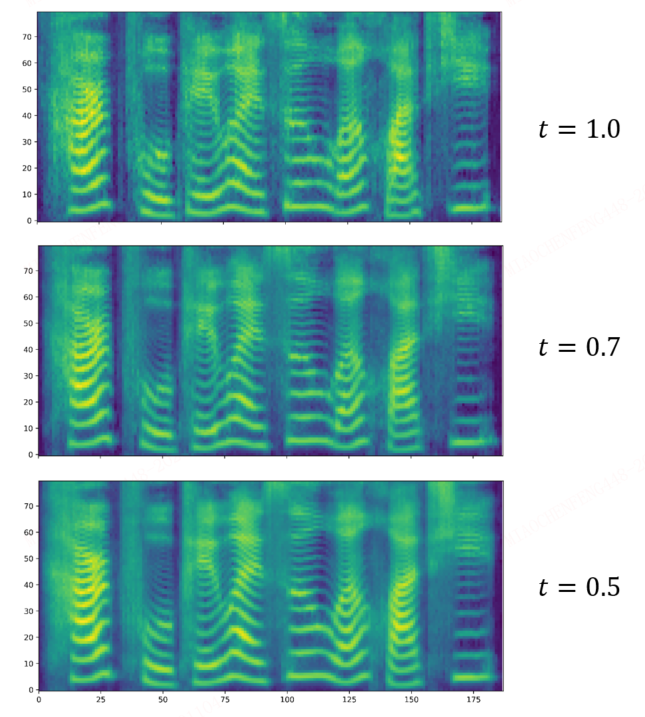}
    \caption{ Melspectrograms with different speech variations generated by EFTS-Flow.
    }\label{variations}
 \end{figure}

\textbf{Speech variation}
 EFTS-Flow is a flow-based model, which is able to produce speech in diversity by input with different latent variable $z$. Fig. \ref{variations} shows the different melspectrograms by changing the temperature of $z$. our results show that EFTS-Flow can produce high quality speech with different variations.
\section{Experiments on Tacotron 2}
\subsection{Implementation of Tacotron2-SMA and Tacotron2-HMA}
We implement Tacotron2-SMA by adding a SMA loss during training.  
The detailed implementation of Tacotron2-HMA is shown in Alg. \ref{algorithm_tacotron2}. For each output timestep, we compute IMV $\pi$ following Eq. (\ref{defini}). A $Clamp(\cdot)$ operation is used to limit $0 \le \Delta \pi \le 1$. We reconstruct the alignment $\alpha$ according to Eq. (\ref{recons2}), and use the newly generated $\alpha$ for further computation.

\begin{algorithm}[tb]
    \caption{Tacotron2 with HMA}
    \label{algorithm_tacotron2}
 \begin{algorithmic}
    \STATE {\bfseries Initialize:} 
        \STATE \ \ \ \ $ \pi = 0$
        \STATE \ \ \ \ $ p = range(0,T_1)$        
    \FOR{$i=0$ {\bfseries to} $T_2-1$}
    \STATE $ \alpha_i = SoftAttention(x_i,h)$
    \STATE $ \pi' = \alpha_i \cdot p$
    \STATE $ \Delta \pi = Clamp(\pi' - \pi, min=0, max=1)$
    \STATE $  \pi  = \pi + \Delta \pi$
    \STATE $\alpha_i = softmax(-\sigma^{-2}*(\pi-p)^2) $
    \STATE $out = DecoderRNN(\alpha_i,h)$
    \ENDFOR
 \end{algorithmic}
 \end{algorithm}
 
 \subsection{Experiments results}
\begin{figure}   
    \centering
    \includegraphics[width=14.2cm,height=9.6cm]{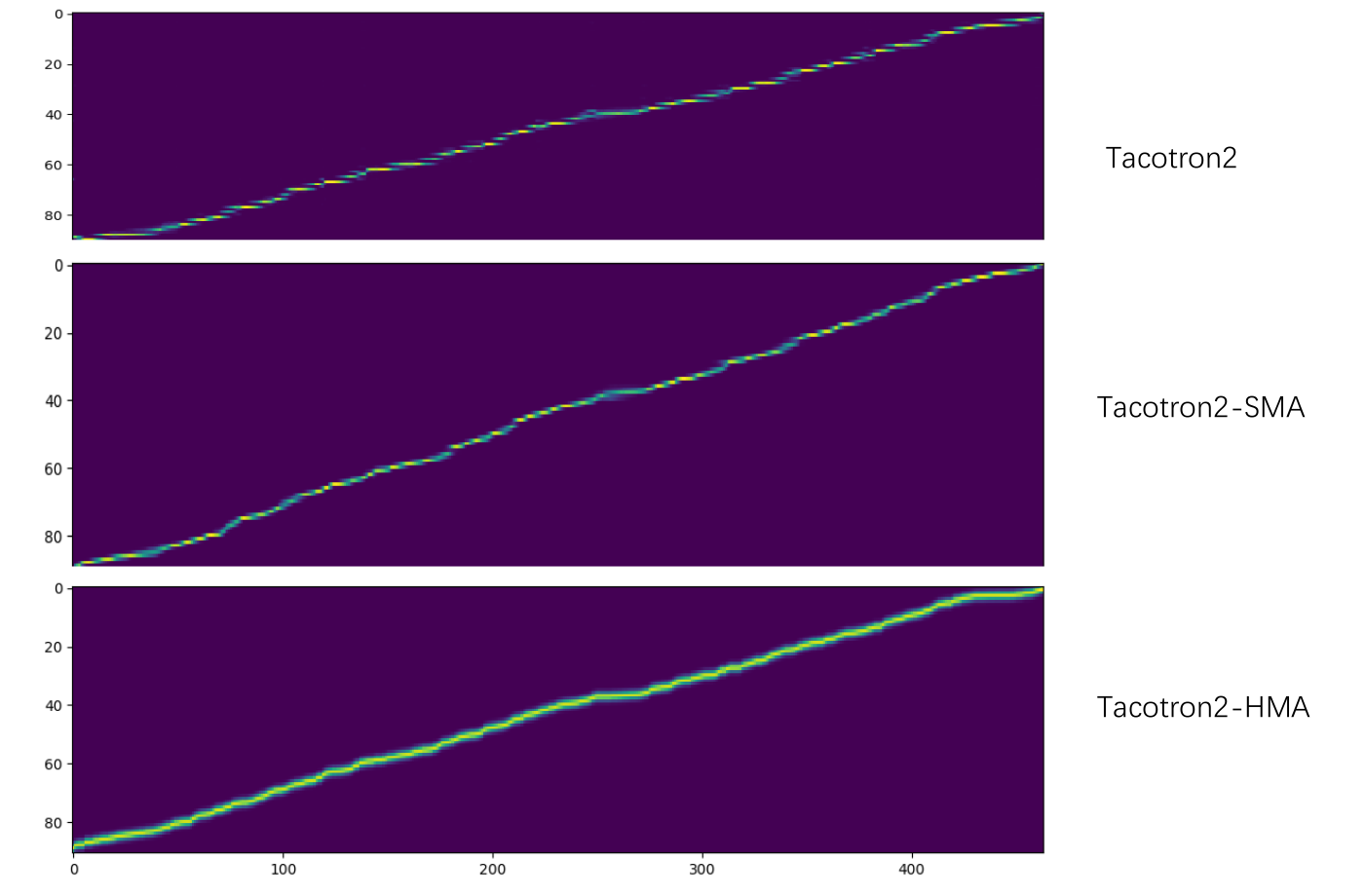}
    \caption{ Alignments of Tacotron 2 with different settings of the same utterance.}\label{att}
 \end{figure}

 We visualized the alignment matrix of different models in Fig. \ref{att}. As can be seen, The alignment matrix of Tacotron 2 is noisy, which often leads to mispronunciations, in contrast, both Tacotron2-SMA and Tacotron2-HMA learn a clean and smooth alignment.
\section{EfficientTTS Pseudocode}
\begin{figure*}[t]
    \centering
   
    \subfigure{}{
    \begin{minipage}[t]{.99\linewidth}
    \includegraphics[width=13cm,height=9cm]{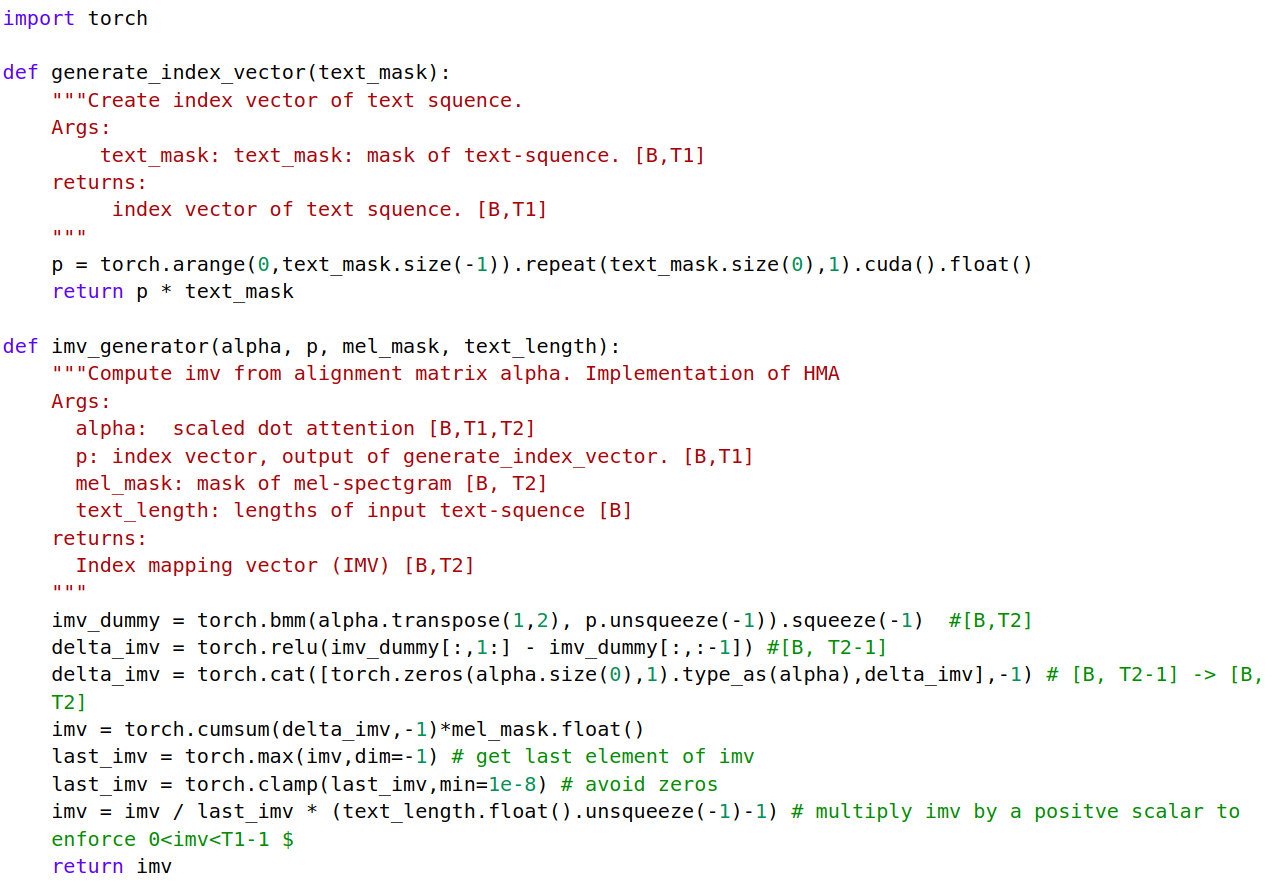}
    \centering
    \end{minipage}
    }
    \subfigure{}{
    \begin{minipage}[t]{.99\linewidth}
    \includegraphics[width=13cm,height=4.5cm]{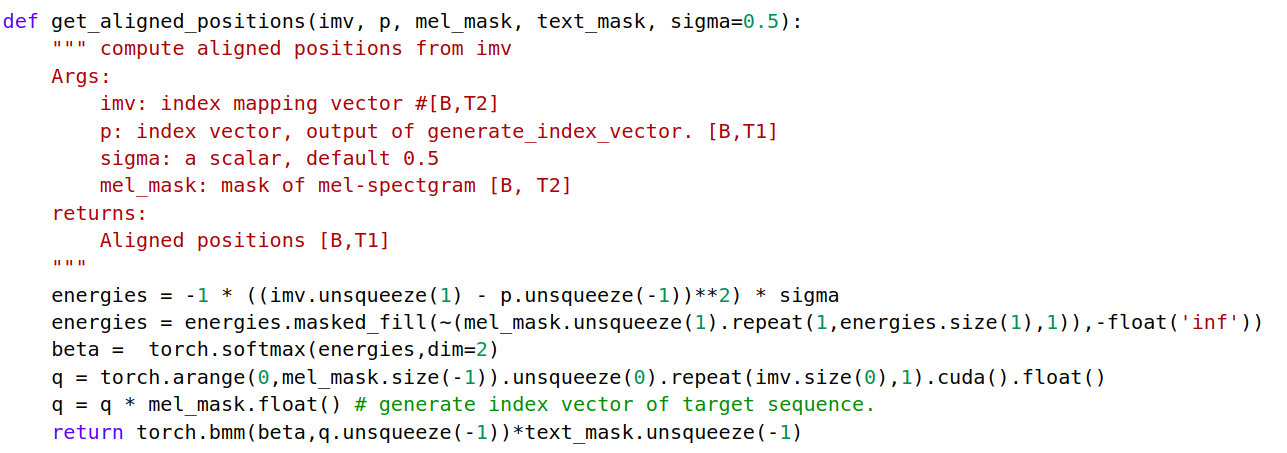}
    \centering
    \end{minipage}
    }
    \subfigure{}{
    \begin{minipage}[t]{.99\linewidth}
    \includegraphics[width=13cm,height=5.9cm]{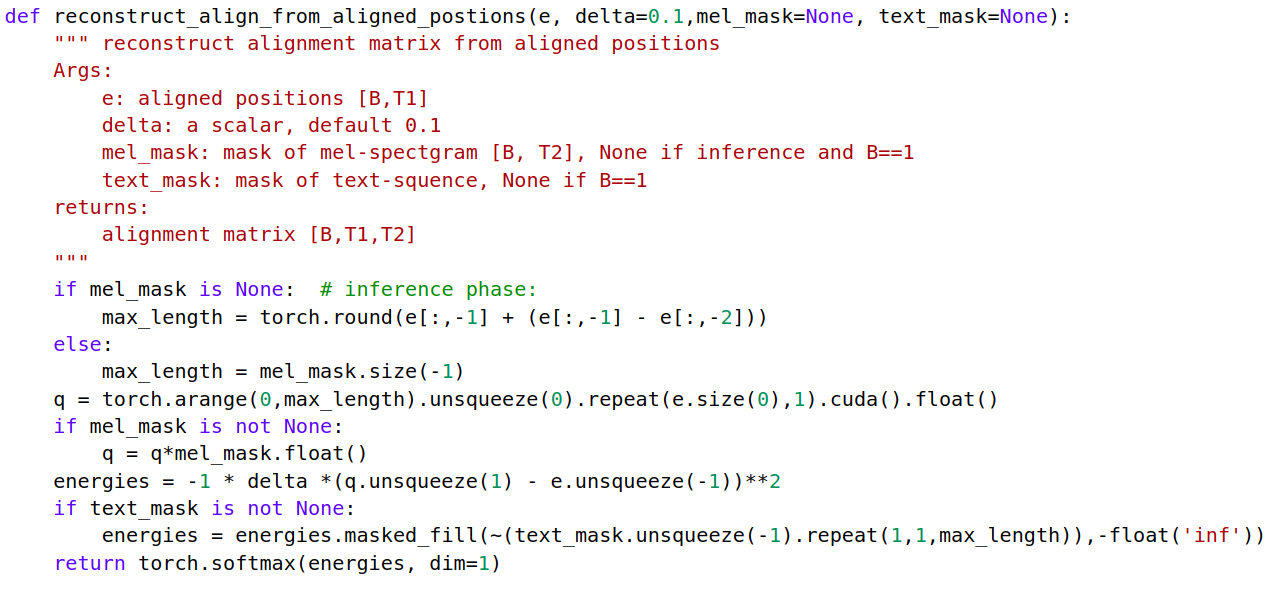}
    \centering
    \end{minipage}
    }
    \caption{EfficientTTS pseudocode } \label{code_fig}

  \end{figure*}

In Fig. \ref{code_fig} we present the pseudocode of EfficientTTS, including generating IMV from alignment matrix, extracting aligned positions from IMV, and reconstructing alignment matrix from aligned positions.

\end{document}